\documentclass[11pt,twoside]{article}

\usepackage{asp2006}
\usepackage{epsf}
\usepackage{psfig}
\usepackage{lscape}
\def\HI{H{\,\small I}}
\newcommand{\kms}{$\,$km$\,$s$^{-1}$}

\begin{document}
\title{Fast neutral outflows in nearby radio galaxies: a major source of feedback}   
\author{Raffaella Morganti$^{1,2}$, Clive Tadhunter$^{3}$, Tom
Oosterloo$^{1,2}$, \\
Joanna Holt$^3$, Bjorn Emonts$^{4}$}

\affil{1) Netherlands Foundation for Research in Astronomy, Postbus 2, NL-7990
AA, Dwingeloo, The Netherlands} 
\affil{2) Kapteyn Astronomical Institute,
University of Groningen, P.O.  Box 800, 9700 AV Groningen, the Netherlands}
\affil{3) Dep.  Physics and Astronomy, University of Sheffield, Sheffield, S7
3RH, UK} 
\affil{4) Dep. of Astronomy, Columbia University, 550 W 120th Street,  NY
10027, USA}

\begin{abstract}
Fast ($\sim 1000$ \kms) outflows of neutral gas (from 21-cm \HI\ absorption)
are detected in strong radio sources.  The outflows occur, at least in some
cases, at distances from the radio core that range between few hundred parsecs and
kpc.  These \HI\ outflows likely originate from the interaction between radio
jets and the dense surrounding medium.  The estimated mass outflow rates are
comparable to those of moderate starburst-driven superwinds.  The impact on
the evolution of the host galaxies is discussed.
\end{abstract}
\section{Introduction}   
Huge amounts of energy are produced through the accretion of material on to
the super-massive black hole situated in the centre of an Active Galactic
Nucleus (AGN).  This energy is released into the surrounding medium in a
number of different ways, ranging from collimated radio-plasma jets to UV and
X-ray emission.  The regions around an AGN are, therefore, highly complex and
host a wealth of physical processes.  Gas in different phases (atomic,
molecular and ionised) is observed in this very hostile environment. This gas
- in particular its kinematics and ionisation - can bear the
signature of the effects of the AGN on its surrounding medium.For example,
the energy released from the nucleus can produce gas outflows of very high
velocities (thousands \kms).  

Gas outflows have a wide range of effects, including clearing the
circum-nuclear regions and halting the growth of the supermassive black-holes
(see e.g.  Silk \& Rees 1998, Di Matteo et al.\ 2005).  These effects can be
at the origin of correlations found between the masses of the central
super-massive black-hole and the properties of the host galaxies.  They can
also prevent the formation of too many massive galaxies in the early
universe. In addition, outflows can inject energy and metals into the
interstellar and intergalactic medium.  It has been suggested that AGN-driven
outflows are a major source of feedback in the overall galaxy formation
process.

Outflows of {\sl ionised} gas have been observed in many AGN, from Seyfert
galaxies to quasars (see e.g.  Crenshaw, Kraemer \& George 2003 and Crenshaw
these proceedings).  In addition to this, we have recently discovered fast and
massive outflows of {\sl neutral hydrogen} detected as 21-cm \HI\ absorption
against the central regions of radio-loud galaxies (Morganti, Tadhunter \&
Oosterloo 2005).  The characteristics and the possible effects of these
outflows are summarised in this contribution.

\section{Fast \HI\ outflows} 

Using the Westerbork Synthesis Radio Telescope (WSRT) we have detected very broad
\HI\ absorption components in 6 radio galaxies.  Three examples of the detected
\HI\ absorption are shown in Fig. 1. 
The broad absorption is observed against the strong
nuclear radio continum.  This broad \HI\ component has been also found in a radio-loud Seyfert
galaxy (IC~5063) observed with the Australia Telescope Compact Array (ATCA).  These
objects are characterised by the presence of a rich ISM surrounding the AGN,
e.g. strong CO or far-IR emission, and/or known to have undergone a major
episode of star formation in the recent past (Tadhunter et al.\ 2005).
Some of them have strong,
steep-spectrum core emission (on a scale $<$10 kpc, i.e. unresolved at the
resolution of the WSRT 21-cm observations). These objects are considered to be
young or recently restarted radio sources.  In many of the objects the presence of a deep {\sl but narrow} \HI\
component  was
already known and studied, but the {\sl shallow broad} component was
discovered only by the new observations  (thanks to the broad band that the
upgraded WSRT could offered).

The broad \HI\ component is typically very shallow (peak optical depth as low
as $\tau \sim 0.0005$), with the full width at zero intensity (FWZI) of the
absorption ranging between 600 \kms\ and almost 2000 \kms\ --- perhaps the
broadest found so far in \HI\ 21-cm absorption in any class of astronomical
objects. A large fraction of the \HI\ absorption is blueshifted relative to
the systemic velocity of the galaxy. Because the gas producing the absorption
must be in front of the radio source, this provides unambiguous evidence that
the bulk of gas is outflowing.  The typical column densities found for the
broad absorption are 1 to 10$\times 10^{21}$ cm$^{-2}$ (for a $T_{spin} =
1000$ K). More details can be found in Morganti et al. (2003, 2005a,b), Emonts
et al. (2003).

In order to understand the origin of such outflows and quantify their effects
it is crucial to know their location as well as obtain a complete as possible
view of what are the characteristics of the gas in other phases
(e.g. ionised gas).

\section{What is the origin?}

Understanding the driving mechanism(s) of the outflows is crucial for
understanding more about the physical mechanisms at work in the central
regions of AGN.  Radiation or wind pressure from the regions near the active
super-massive black hole (i.e.  a quasar wind) are the likely drivers of the
gas outflows detected in X-ray and UV.  However, in radio-loud objects, an
other possible mechanism for driving the outflows is the interaction of the
radio plasma with the (rich) gaseous medium in the direct vicinity of the
active nucleus (Tadhunter 2007). In the case of the fast \HI\ outflows
presented here, there is now strong evidence that this is likely to be the
case.  This evidence comprises follow-up observations at higher resolution of
two of the objects in our sample (IC~5063 and 3C~305, Oosterloo et al. 2000
and Morganti et al. 2005 respectively) that demonstrate that the outflow
regions are resolved on a scale of 200~pc (IC~5063) and 1.6~kpc (3C~305), and
spatially associated with both the extended, bright  radio lobes and the
outflows of ionised gas detected at the same locations in optical observations.  In
a third case (3C~293), the evidence that the outflow is located at $\sim 1$
kpc is more indirect (Morganti et al. 2003, Emonts et al. 2005). The presence
of neutral atomic gas accelerated to such high velocities is in itself
intriguing.  It indicates that after a strong jet-cloud interaction the gas
can cool very efficiently.  Such rapid cooling is indeed predicted by recent
numerical simulations of jets impacting on gas clouds (Mellema et al.\ 2002,
Fragile et al.\ 2004, Krause 2007).

\begin{figure}
\centerline{\psfig{figure=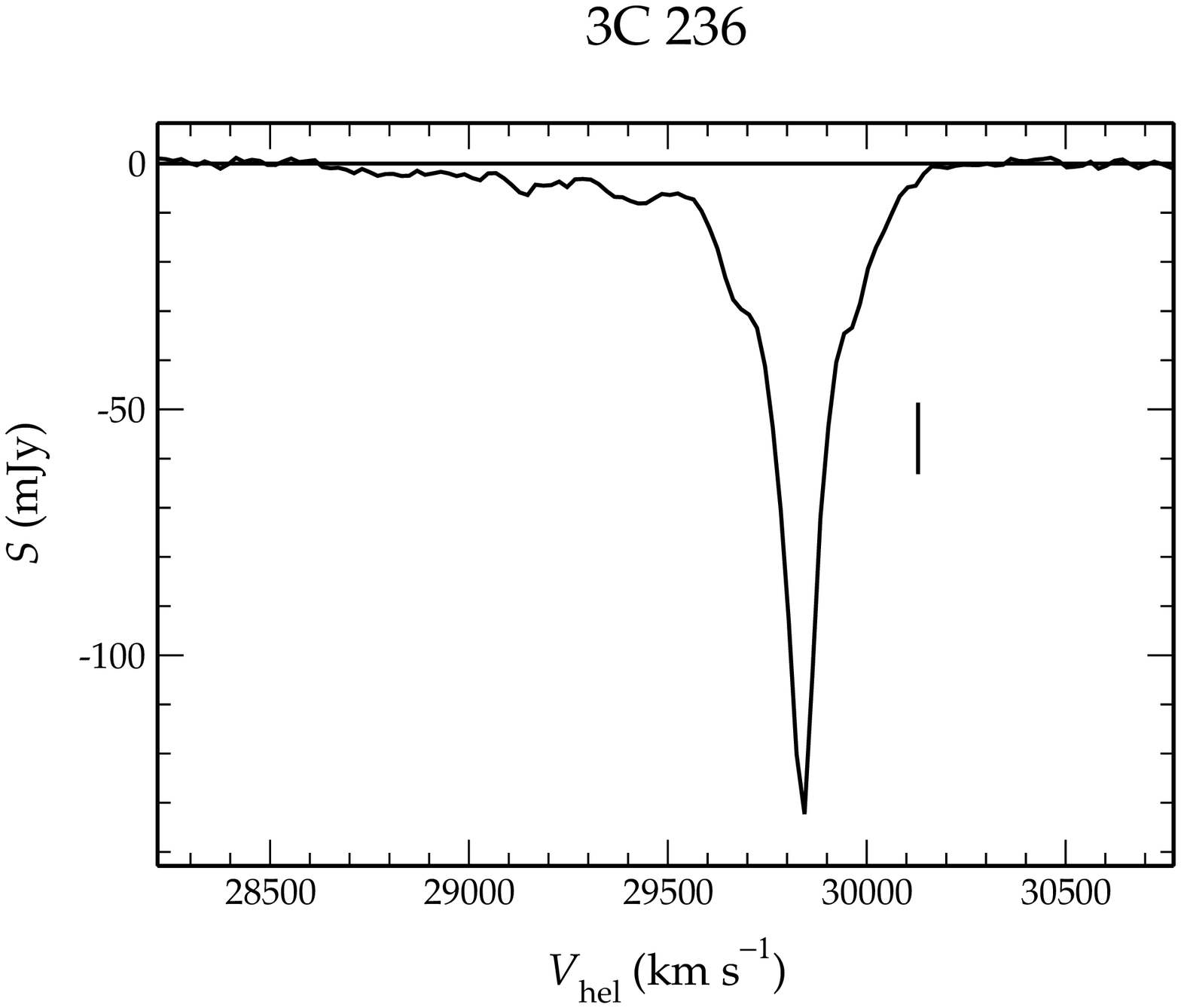,angle=0,width=4.5cm}, 
\psfig{figure=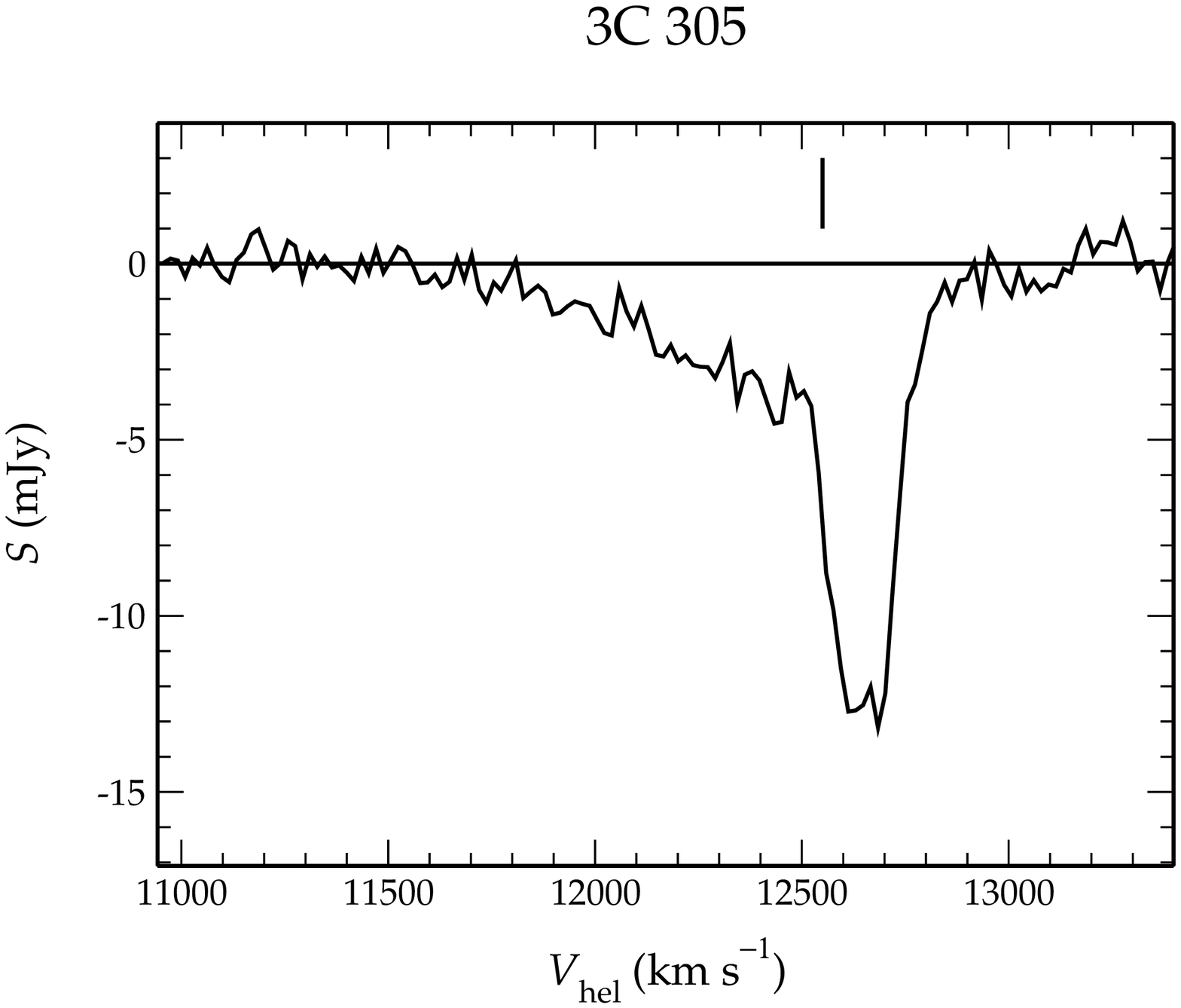,angle=0,width=4.5cm}
\psfig{figure=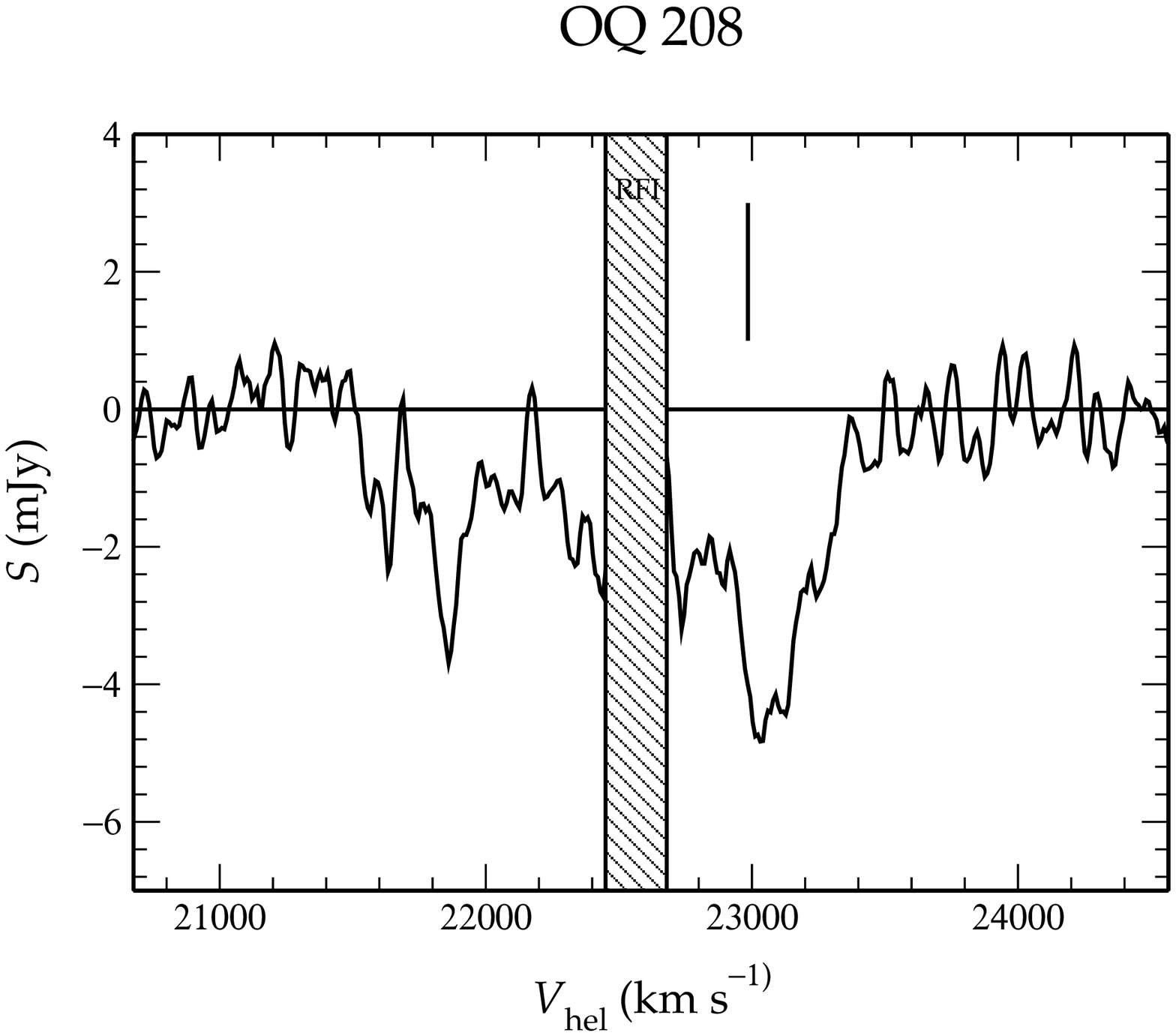,angle=0,width=4.5cm}
}
\caption{21 cm - \HI\ absorption profiles detected against three radio
galaxies. 
The observations of the radio galaxies were done using the upgraded
WSRT. The vertical line indicates the systemic velocity.
}
\end{figure}

\subsection{Effects on the galaxy}

The main result of this study is that the neutral outflows occur, in at least
some cases, at kpc distance from the nucleus, and they are most likely driven
by the interactions between the expanding radio jets and the gaseous medium
enshrouding the central regions.  We estimate that the associated mass outflow
rates are up to $\sim 50$ $M_\odot$ yr$^{-1}$, comparable (although at the
lower end of the distribution) to the outflow rates found for starburst-driven
superwinds in Ultra Luminous IR Galaxies (ULIRG), see Rupke et al. (2002).
This suggests that massive, jet-driven outflows of neutral gas in radio-loud
AGN can have as large an impact on the evolution of the host galaxies as the
outflows associated with starbursts. This is important as starburst-driven
winds are recognised to be responsible for inhibiting early star formation,
enriching the ICM with metals and heating the ISM/IGM medium.

In few cases we have done a more detailed study of the characteristics of the
outflow to investigate whether its properties are consistent with the
assumption of the AGN feedback models.  Particularly interesting is the case
of IC~5063, where the mass outflow rates of cold (\HI) and warm (ionised) gas
have been found to be comparable, $\sim 30$ $M_{\odot}$ yr$^{-1}$.  With a
black-hole mass of $2.8 \times 10^8$ M$_\odot$ (Nicastro, Martochia \& Matt
2003), the Eddington luminosity of IC~5063 is $3.8 \times 10^{46}$ erg
s$^{-1}$, this means that the kinetic power of the outflow represents only
about few $\times 10^{-4}$ of the available accretion power (Morganti et
al. in prep).  This result is similar to what found for PKS~1549-79 (Holt et
al.  2006 and these proceedings).  However, unlike the case of PKS~1549-79,
IC~5063 accrete at low rate ($\dot m \sim 0.02$, Nicastro et al.
2003). Thus, in IC~5063 the kinetic power of the outflow appears to be a
relative high fraction of the nuclear bolometric luminosity ($\sim 8 \times
10^{-2}$) compared to what found in some more luminous radio galaxies. This
might be relevant for the feedback process.

\begin{figure}
\centerline{\psfig{figure=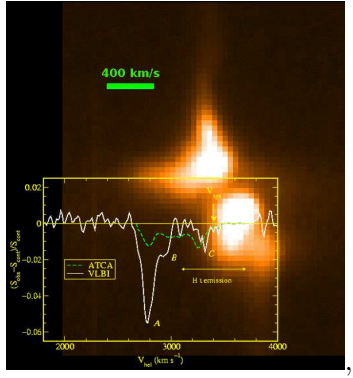,angle=0,width=4.5cm}, 
}
\caption{Comparison between the width of the \HI\ absorption (white
  profile, Oosterloo et al. 2000) and that of the ionised gas (from the {[O
  III]}5007\AA). The first order similarity between the amplitude of the
  blueshifted component covered by the two phases of the gas is clearly seen.
  }
\end{figure}

\end{document}